\documentclass[%
aip, cha,%aps, prl,
floatfix,
preprintnumbers,
reprint,
 superscriptaddress,
amsmath, amssymb,
onecolumn,
]{revtex4-1}

\usepackage[utf8]{inputenc}%for special characters
\usepackage{graphicx}% Include figure files
\usepackage{bm}% bold math
\usepackage{color}
\usepackage{amsmath,amssymb,amstext,amsthm,relsize}
\usepackage[breaklinks=true]{hyperref}
\hypersetup{bookmarksnumbered, pdfpagemode=UseOutlines, 
	pdfauthor={Ivan J.\ Vera-Marun}, 
	pdftitle={Independent geometrical control of spin and charge resistances in curved spintronics}, 
	pdfdisplaydoctitle, 
	colorlinks=true, citecolor=blue, filecolor=blue, linkcolor=blue, urlcolor=blue}

\begin{document}
	
\title{Independent geometrical control of spin and charge resistances in curved spintronics}
	
	\author{Kumar Sourav \surname{Das}}
	\email[e-mail: ]{K.S.Das@rug.nl}
\affiliation{Physics of Nanodevices, Zernike Institute for Advanced Materials, University of Groningen, 9747 AG Groningen, The Netherlands}
	%\homepage[]{} \thanks{}
	\author{Denys Makarov}
	\affiliation{Helmholtz-Zentrum Dresden-Rossendorf e.V., Institute of Ion Beam Physics and Materials Research, Bautzner Landstrasse 400, 01328 Dresden, Germany}
	\author{Paola Gentile}
%	\affiliation{CNR-SPIN and Dipartimento di Fisica “E. R. Caianiello”, Università degli Studi di Salerno, I-84084 Fisciano, Italy}
	\affiliation{CNR-SPIN, c/o Università degli Studi di Salerno, I-84084 Fisciano, Italy}
	\affiliation{Dipartimento di Fisica “E. R. Caianiello”, Università degli Studi di Salerno, I-84084 Fisciano, Italy}
	\author{Mario Cuoco}
	\affiliation{CNR-SPIN, c/o Università degli Studi di Salerno, I-84084 Fisciano, Italy}
	\affiliation{Dipartimento di Fisica “E. R. Caianiello”, Università degli Studi di Salerno, I-84084 Fisciano, Italy}
	\author{Bart J.\ \surname{van Wees}}
\affiliation{Physics of Nanodevices, Zernike Institute for Advanced Materials, University of Groningen, 9747 AG Groningen, The Netherlands}
	\author{Carmine Ortix}
	\email[e-mail: ]{C.Ortix@uu.nl}
	\affiliation{{Institute for Theoretical Physics, Center for Extreme Matter and Emergent Phenomena, Utrecht University, Princetonplein 5, 3584 CC, Utrecht, The Netherlands}}
	\affiliation{Institute for Theoretical Solid State Physics, IFW Dresden, 01171 Dresden, Germany}
	\affiliation{Dipartimento di Fisica “E. R. Caianiello”, Università degli Studi di Salerno, I-84084 Fisciano, Italy}
	\author{Ivan J.\ Vera-Marun}
	\email[e-mail: ]{ivan.veramarun@manchester.ac.uk}
	\affiliation{School of Physics and Astronomy, University of Manchester, Manchester M13 9PL, United Kingdom}

	
	%\preprint{Draft 22}
	
\begin{abstract}

\textbf{Spintronic devices operating with pure spin currents represent a new paradigm in nanoelectronics, with higher energy efficiency and lower dissipation as compared to charge currents. %\cite{jedema_electrical_2001, valenzuela_spin-polarized_2004, kimura_large_2007, kimura_temperature_2008, lou_electrical_2007, van_t_erve_electrical_2007, tombros_electronic_2007}.
This technology, however, will be viable only if the amount of spin current diffusing in a nanochannel can be tuned on demand while guaranteeing electrical compatibility with other device elements, to which it should be integrated in high-density three-dimensional architectures. %\cite{m._sekikawa_novel_2008, parkin_magnetic_2008, parkin_memory_2015, lavrijsen_magnetic_2013}.
Here, we address these two crucial milestones and demonstrate that pure spin currents can effectively propagate in metallic nanochannels with a three-dimensional curved geometry. Remarkably, the geometric design of the nanochannels can be used to reach an independent tuning of spin transport and charge transport characteristics. These results put the foundation for the design of efficient pure spin current based electronics, %\cite{zutic_spintronics:_2004, schmidt_fundamental_2000}
which can be integrated in complex three-dimensional architectures.} 
	
\end{abstract}

\maketitle
	
A number of next-generation electronic devices, including memory elements and transistor circuits, rely on spin currents. Pure spin currents \cite{jedema_electrical_2001, valenzuela_spin-polarized_2004, kimura_large_2007, kimura_temperature_2008, lou_electrical_2007, van_t_erve_electrical_2007, tombros_electronic_2007} transfer only spin angular momentum and therefore have the additional advantage that the electronic devices can operate with low power dissipation. A pure spin current can be generated 
using the coupling between charge and spin transport across the interface of a ferromagnet with a contiguous paramagnetic nanochannel. 
The efficiency of the spin injection across this interface can be optimized by improving the interface quality and the device structure. The propagation of the pure spin current along the nanochannel is instead related to its spin relaxation length. In conventional metals and small-gap semiconductors, the dominant spin relaxation mechanism corresponds to the so-called Elliot-Yafet mechanism  \cite{kimura_temperature_2008,zutic_spintronics:_2004,jedema_spin_2003}, which dictates that the spin relaxation length is strictly locked to the resistivity of the metallic paramagnet. 
%As a result, any control on the amount of transported pure spin current, which would be highly desirable for spintronic applications, is inevitably reflected in a change in the electrical resistance. 
This, in turn, severely compromises the applicability of pure spin currents to technologically relevant modern electronics, which necessitates the individual matching of spin and charge resistances in order to achieve efficient coupling of spin and charge degrees of freedom \cite{schmidt_fundamental_2000, rashba_theory_2000, zutic_spintronics:_2004}. 

\begin{figure*}[tbp]
		\includegraphics*[angle=0, trim=0mm 0mm 0mm 0mm, width=170mm]{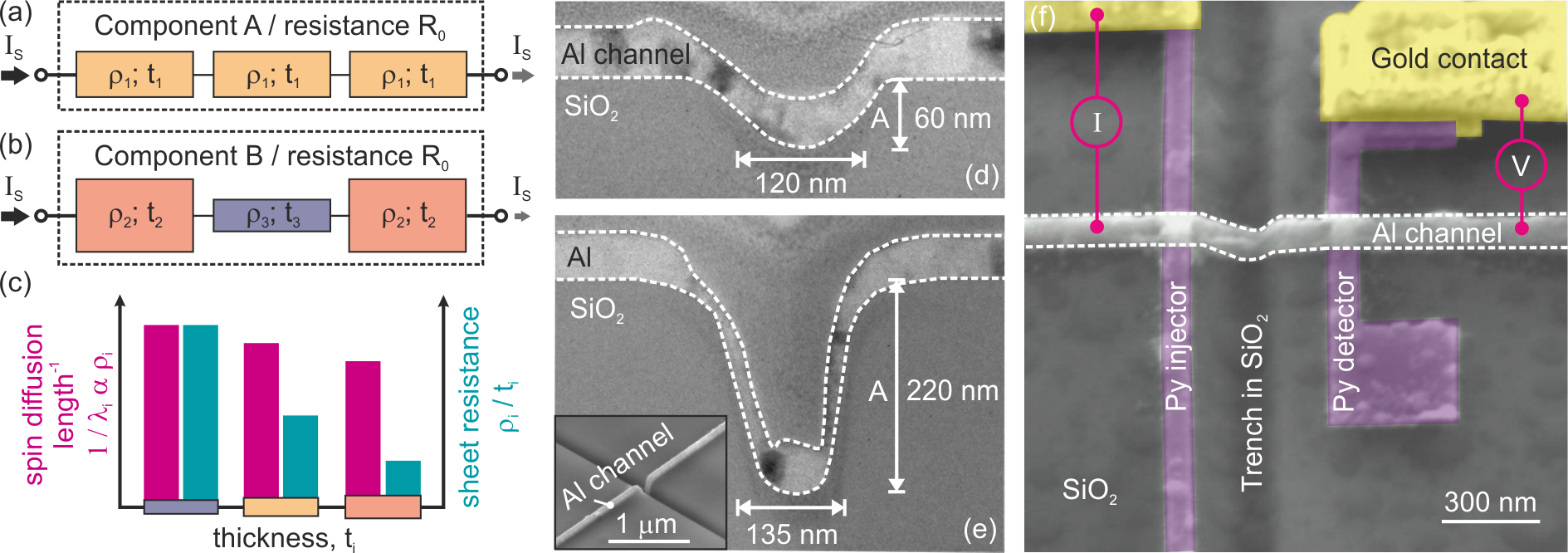}
		\caption{
			\label{fig:NLSV-SEM}
			\textbf{Concept of geometrical control of spin current and curved device architecture.} \textbf{(a-b)} Schematics of two different spin transport channels, each composed of three elements in series. The  elements of the channel in (a) are identical, representing a homogeneous channel, resulting in a total charge resistance $R_0$ and a spin current $I_\text{s}$. The channel in (b) is inhomogeneous, with components having different thicknesses and resistivities ($\rho$), and still with a total charge resistance $R_0$. However, its spin resistance is differently modulated with the thickness, resulting in a different spin current as compared to the homogeneous channel in (a). \textbf{(c)} Distinct role of channel thickness ($t$) on the modulation of sheet resistance $\rho / t$ and of the spin relaxation length ($\lambda$), leading to distinct scaling of charge and spin resistances. \textbf{(d-e)} Transmission electron microscope (TEM) cross-sections of Al channels grown on trenches of different geometries, characterized by the trench height $A$ and the full width at half maximum. Top-view of an Al channel grown across a trench is shown in the scanning electron microscope (SEM) image in the inset of panel (e). \textbf{(f)} SEM image of a spin valve device with a curved Al channel across a trench. The electrical connections for non-local spin valve measurements are also depicted.   
		}
	\end{figure*}

Here, by using a combination of experimental investigations and theoretical analysis, we show that spin and charge resistances can be independently tuned in metallic  nanochannels. Importantly, this is realised 
%without the need of any strong electronic correlations as those studied in interacting 1D systems \cite{giamarchi_quantum_2003}and 
even in the absence of any external electric or magnetic gating \cite{villamor_modulation_2015, dejene_control_2015}, 
and it is totally different in nature to the spin-charge separation phenomenon in Tomonaga-Luttinger liquids \cite{giamarchi_quantum_2003, tsvelik_quantum_2015}. 
Our strategy relies on the possibility to grow metallic nanochannels with a strongly inhomogeneous nanometer-scale thickness, $t$. The size-dependent resistivity, $\rho$, of the metallic channels \cite{steinhogl_size-dependent_2002} yields a different local behaviour for the sheet resistance $\rho/t$ and the spin relaxation length $\lambda \propto 1/\rho$  [c.f.\ Fig.~\ref{fig:NLSV-SEM}(a-c)]. As a result, an appropriate engineering of the nanochannel thickness allows to design nanochannels where one can achieve independent tuning of spin resistance without affecting the total charge resistance, and {\it vice versa}. This capability allows for the design of an element with simultaneous matching of spin resistance to a spin-based circuit, e.g.\ for efficient spin injection \cite{schmidt_fundamental_2000, rashba_theory_2000, zutic_spintronics:_2004}, and matching of charge resistance to a charge-based circuit, e.g.\ for efficient power transfer. 
The control of spin and charge resistances is fundamental to spintronics, as it enables practical magnetoresistance in two terminal devices \cite{fert_conditions_2001} and the concatenability and reduced feedback in spin logic architectures \cite{dery_spin-based_2007, behin-aein_proposal_2010}.

As a proof of concept, we demonstrate modulation of spin currents and of charge currents in lateral non-local spin valves \cite{jedema_electrical_2001} with ultrathin metallic channels directly grown on curved templates [c.f.\ Fig.~\ref{fig:NLSV-SEM}(d,e)], thereby allowing us to achieve efficient spin current propagation in three-dimensional nanoarchitectures. 
This is of immediate relevance when considering a practical implementation of spintronics. 
On the one hand, transport of pure spin currents in non-local spin valves is at the heart of multiple proposals of spin-based logic architectures \cite{dery_spin-based_2007, behin-aein_proposal_2010}, and thus of potential technological impact. 
On the other hand, the use of curvature to independently control spin and charge impedances in multi-terminal devices adds a novel approach for their efficient integration with complementary metal oxide semiconductor (CMOS) transistors that optimizes device reliability and endurance \cite{makarov_cmos-compatible_2016}. 
Finally, as CMOS technology scales down to 10~nm features or less, there are increasing efforts in the development of three-dimensional CMOS microelectronics that can overcome the limitations of Moore's law. This is similar with regards to spintronics and its integration with CMOS. Such efforts have been led by the thinning and 3D stacking of several chips, initially integrating CMOS and spin-based memories \cite{m._sekikawa_novel_2008} and later extended to heterogeneous chips \cite{m._koyanagi_heterogeneous_2013}. 
A completely different approach is to change the architecture itself to be three-dimensional. Until now, the realisation of vertical flow of spin information via three-dimensional channels has been based solely on the movement of magnetic domain walls, by applying current \cite{parkin_magnetic_2008} or magnetic field \cite{lavrijsen_magnetic_2013}, with a recent implementation based on depositing magnetic material on the side-wall of deep trenches \cite{parkin_memory_2015}. 
Our work on curved nanoarchitectures for pure spin current devices delves into territory so far only explored for charge-based technologies. 
While being conceptually simple and potentially cheap, it offers the possibility of high density three-dimensional integration over that in conventional spin current devices.

% \subsection*{Sample fabrication}
Curved templates were created in the form of trenches in a silicon dioxide substrate. Increasing the height of the trenches $A$ [c.f.\ Fig.~\ref{fig:NLSV-SEM}(d,e)] led to channels with increasing curvature, allowing us to systematically explore the effect of channel geometry. 
To create the trenches we used focused ion beam (FIB) etching, where the geometry of the trenches was controlled by varying the FIB milling times.
%, leading to different height and curvatures, as shown in Fig.~\ref{fig:NLSV-SEM}(d-e). In order to categorize the different trench geometries we use the trench height, $A$, extracted from transmission electron microscope images. 
Each sample consists of two lateral spin valve devices: one device with the spin transport channel across the trench, resulting in a curved device, and another on the flat part of the substrate, serving as a reference device. The spin valve devices were prepared by multi-step e-beam lithography, e-beam evaporation of materials and resist lift-off techniques, as described in Ref.~\onlinecite{das_anisotropic_2016}. Permalloy (Ni$_{80}$Fe$_{20}$, Py) nanowires, with a thickness of 20~nm, were used as the ferromagnetic electrodes. Injector and detector Py electrodes were designed with different widths (80~nm and 100~nm) to achieve different coercive fields. 
	The injector-detector in-plane separation ($L$) was 500~nm for all the devices, except for the one with the largest trench height ($A=270$~nm) which had a separation of $700$~nm.	
	For the spin transport channel we used an aluminium (Al) nanowire, with a width of 100~nm and a nominal thickness of 50~nm. The Al channel was evaporated following a short in-situ ion milling step to remove surface oxide and resist contamination from the Py electrodes, resulting in Al/Py ohmic contacts with a resistance-area product lower than $10^{-15}$~$\Omega.\text{m}^2$. 
	%The separation between the Py injector and detector electrodes was 500~nm for all devices, unless otherwise specified. 
	Fig.~\ref{fig:NLSV-SEM}(f) shows a scanning electron microscope image of one of the fabricated curved spin valve devices.

%\subsection*{Electrical characterization}
All electrical measurements were performed with the sample in a high vacuum environment, within a liquid helium cryostat. The electrical resistance of the Al channel was measured by the four-probe method, with the current applied between the two ends of the Al channel and the voltage drop measured between the injector and the detector electrodes. For the non-local spin valve measurements, the electrical connections are schematically shown in Fig.~\ref{fig:NLSV-SEM}(f). Here, an alternating current ($I$) source, with a magnitude of 400~$\mu$A and frequency of 13~Hz, was connected between the injector electrode and the left end of the Al channel. The non-local voltage ($V$) at the detector electrode, with reference to the right end of the Al channel, was measured by a phase sensitive lock-in technique. 
A magnetic field was applied along the length of the Py wires during these measurements to configure the injector and detector electrodes in a parallel (P) or an anti-parallel (AP) state, corresponding to two distinct levels of the non-local resistance ($R_{\text{NL}}=V/I$). 
The spin valve signal ($\Delta R_{\text{NL}}$) is then given by the difference of the non-local resistance between parallel and anti-parallel configurations, $\Delta R_{\text{NL}}=R_{\text{NL}}^{\text{P}}-R_{\text{NL}}^{\text{AP}}$.
The measurements were carried out at room temperature and at 4.2~K to study spin transport in channels with increasing curvature.
The extraction of $\Delta R_{\text{NL}}$ via this standard low-frequency first-harmonic lock-in technique serves to accurately extract the pure spin current signal and exclude any role of induction or thermoelectric effects \cite{jedema_electrical_2001, das_anisotropic_2016}. 

\begin{figure}[tbp]
	\includegraphics*[angle=0, trim=0mm 0mm 0mm 0mm, width=85mm]{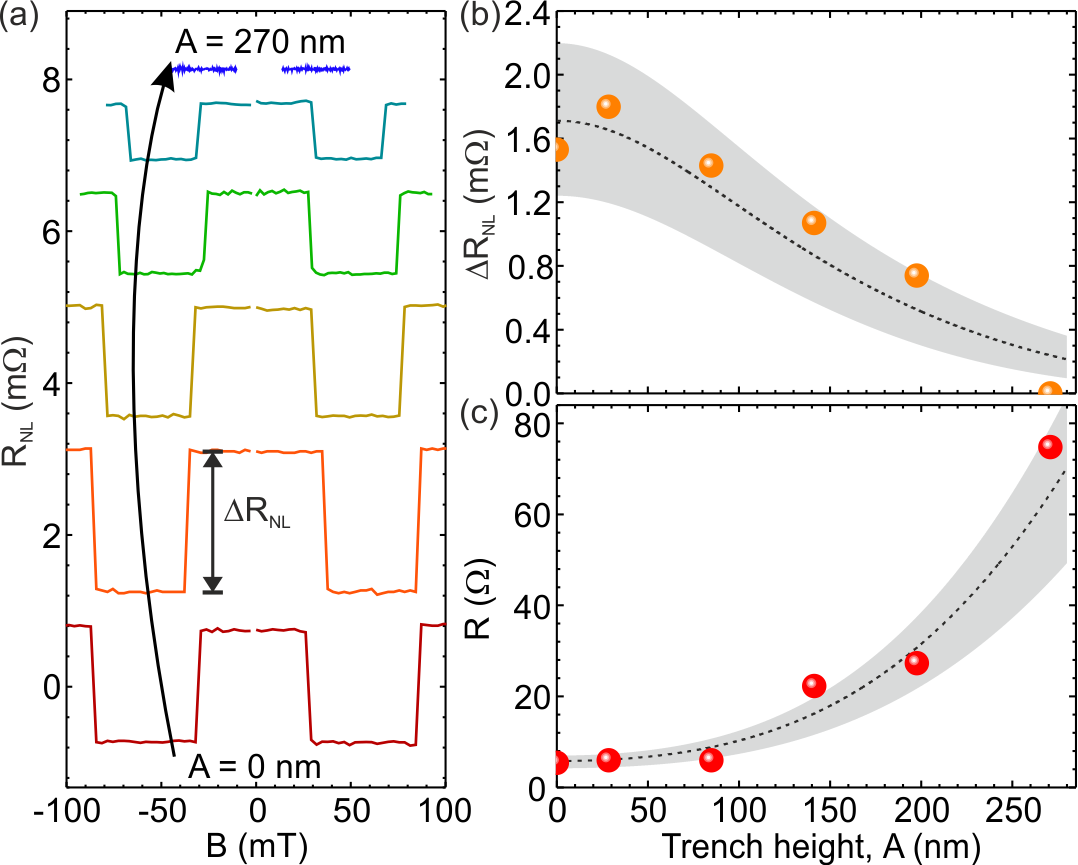}
	\caption{
		\label{fig:Experiment}
		\textbf{Non-local spin valve signal and channel resistance measurements and modelling.} \textbf{(a)} Spin valve measurements at $T=4.2$~K for devices with different channel geometries. The black arrow indicates the direction of increasing trench height, $A$. The spin signal $\Delta R_{\text{NL}}$ decreases with increasing $A$. \textbf{(b)} $\Delta R_{\text{NL}}$ as a function of $A$. The experimental data and the modelling result are shown as solid spheres and dotted line, respectively. The shaded region in grey represents the uncertainty due to device to device variation. \textbf{(c)} Experimental data and modelling results for the charge resistance ($R$) of the channel, for different $A$.
	}
\end{figure}

%\section*{Results}
%\subsection*{Non-local spin transport experiments in curved nanochannels}
The non-local spin valve measurements are shown in Fig.~\ref{fig:Experiment}(a). 
The resulting modulation of $\Delta R_{\text{NL}}$ with $A$ is plotted in Fig.~\ref{fig:Experiment}(b). $\Delta R_{\text{NL}}$ is maximum for the reference spin valves with $A=0$ and shows little change for trenches with $A<50$~nm, limited by device to device variation. However, for increasing trench heights above $\approx 100$~nm we observe a strong decrease in $\Delta R_{\text{NL}}$, until it is fully suppressed for the trench with $A=270$~nm. On the other hand, the measured four-probe charge resistance of the curved channel between the injector and detector electrodes exhibited an opposite trend, as observed in Fig.~\ref{fig:Experiment}(c). Here, a steep increase in resistance ($R$) is seen for  trenches with height greater than $\approx 100$~nm. A similar behaviour was observed at room temperature [see Supporting Information Section 1]. 
	
The contrasting behaviours of both the spin valve signal and charge resistance offer direct evidence of the effect of the curved geometry introduced by the trench. We have first checked that both the strong suppression of $\Delta R_{\text{NL}}$ and the steep increase of $R$ with increasing $A$ cannot be explained just by considering the increase in the channel length due to the curved geometry. To properly describe both of these behaviours we have therefore developed a theoretical model which is applicable to devices, where the local channel geometry explicitly impacts on both charge and spin transport properties. 
Here the key ingredient is the consideration of the dominant Elliot-Yafet spin relaxation mechanism. 
The main outcome of this approach is depicted by the dotted lines in Fig.~\ref{fig:Experiment}(b-c), where \textit{quantitative} agreement with the experimental results is achieved. In the following discussion we introduce this theoretical model.
	
\begin{figure}[t]
		\includegraphics*[angle=0, trim=2mm 2mm 7mm 0mm, width=85mm]{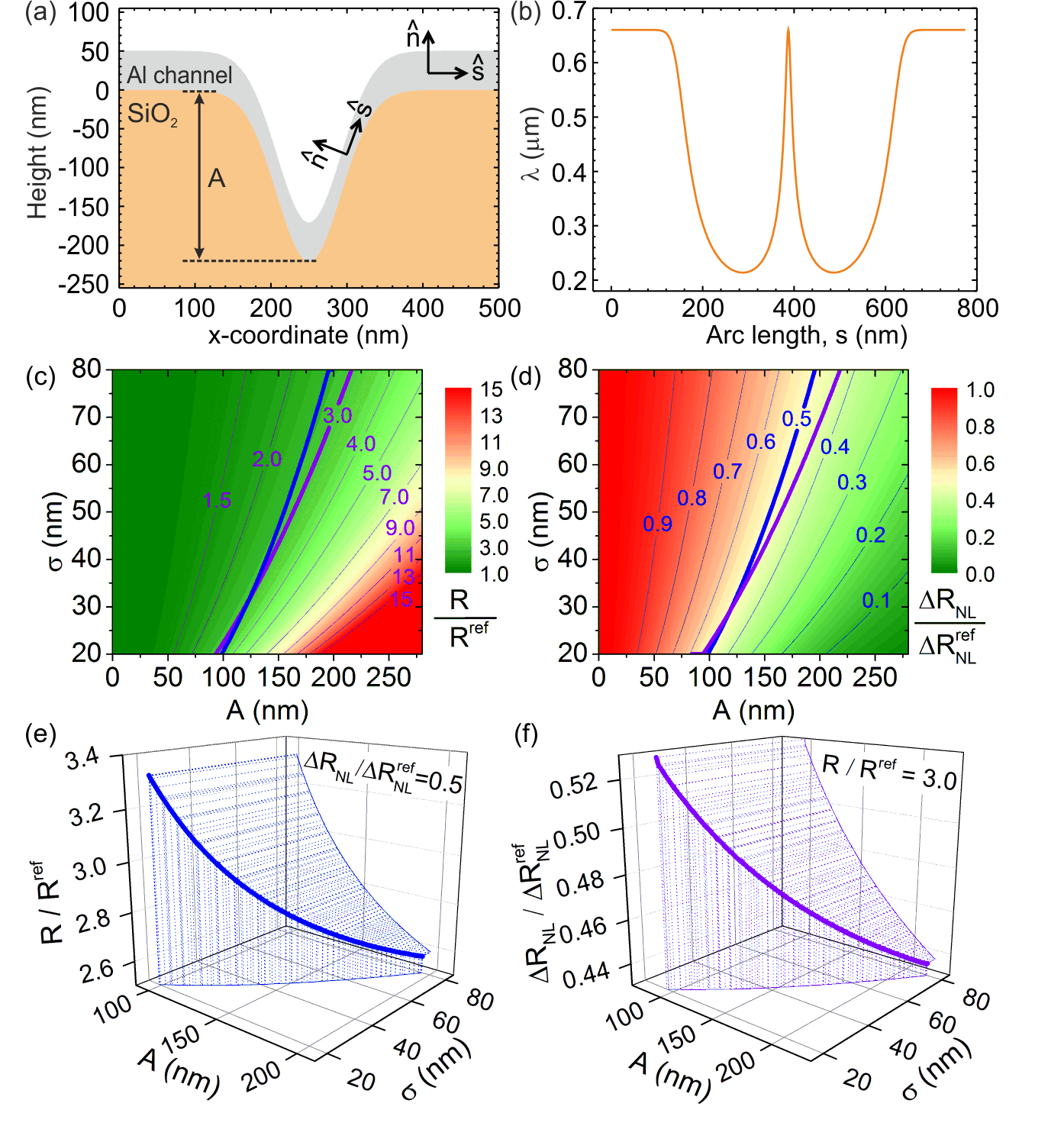}
		\caption{
			\label{fig:contour}
			\textbf{Geometry-induced tuning of charge resistance and spin resistance.} 
			\textbf{(a)} The trench geometry is modelled as a Gaussian bump and the profile of the Al channel across the trench is mapped out. The trench height ($A$) and the unit vector \textit{\^{s}} along the arclength of the Al film, perpendicular to the local surface normal \textit{\^{n}}, have been illustrated.
			\textbf{(b)} Calculated variation of the spin relaxation length in Al along $s$ at 4.2~K.
			\textbf{(c-d)} 2D colour maps illustrating the modulation of charge resistance (c) and spin resistance (d) with the channel geometry, considering a template in the form of a Gaussian bump with height $A$ and full width at half maximum $2\sqrt{2\log{2}}\sigma$ as that in (a). Both the charge ($R$) and the spin ($\Delta R_\text{NL}$) resistances have been normalized by the respective  values for a reference flat channel. A contour line representing $R/R^\text{ref}=3.0$ (thick black) in panel (c) has been projected onto panel (d), and a contour line representing $\Delta R_\text{NL}/\Delta R_\text{NL}^\text{ref}=0.5$ (thick blue) in panel (d) has been projected onto panel (c). \textbf{(e)} 3D plot of the contour line for $\Delta R_\text{NL}/\Delta R_\text{NL}^\text{ref}=0.5$ mapped onto the values of $R/R^\text{ref}$ from panel (c). \textbf{(f)} A similar 3D plot of the contour line representing $R/R^\text{ref}=3.0$ mapped onto the values of $\Delta R_\text{NL}/\Delta R_\text{NL}^\text{ref}$ from panel (d). These results highlight the independent tuning of spin resistance for a constant charge resistance, and {\it vice versa}, via nanoscale design of the template geometry.    
			}
	\end{figure}

To develop an accurate description of the channel we rely on the knowledge of its geometry from TEM imaging (Fig.~\ref{fig:NLSV-SEM}). We observe how at the steep walls of the trench the film thickness was reduced, relative to its nominal thickness. This variation in thickness is determined by the e-beam evaporation technique used to grow the film, where nominal thickness is only achieved when the Al beam impinges on the substrate at normal incidence. With this direct evidence of thickness inhomogeneity, we have incorporated it in our description of the curved channel by modelling the trench profile as a Gaussian bump with FWHM of $\approx 100$~nm, as shown in Fig.~\ref{fig:contour}(a). The resulting thickness of the Al channel in the local surface normal direction \textit{\^{n}} then becomes intrinsically inhomogeneous. 
%\subsection*{Modelling}
We describe 
%model the profile of the trenches created in the silicon dioxide substrates with a 
the Gaussian bump as $h(x)=A \mathrm{e}^{-x^2 / (2 \sigma^2)}$, where the $x$ coordinate is measured with respect to the maximum trench height position. We next consider that the top surface of the evaporated Al film assumes the same profile with $h_{\textrm{T}}(x)= t_0 + h(x)$, and $t_0$ the nominal thickness. With this, the total volume of the evaporated Al channel does not depend on the geometry of the trench, and it is given by $t_0 L w$ where $w$ is the channel width, and $L$ the distance in the $\hat{x}$ coordinate between injector and detector. 
In order to subsequently derive the local thickness profile, we write the line element 
$$d s^2 = \left[1+ \left(\frac{d h(x)}{d x}\right)^2 \right] d x^2, $$ 
which allows to express the arclength measured from the injector electrode as 
\begin{equation}
s(x)= \int_{-L/2}^{x} \sqrt{1 +\left( \frac{d h(x^{\prime})}{d x^{\prime}}\right)^2}.
\label{eq:arclength}
\end{equation}
The channel length between injector and detector is given by $L^{\prime} \equiv s(L/2)$. Furthermore, the local thickness profile can be obtained by requiring 
$\int_{0}^{L^{\prime}} t(s) d s \equiv t_0 L.$ This relation is satisfied for a local thickness profile, which, in terms of the $x$ coordinate can be expressed as 
$$t(x) = \dfrac{t_0}{\sqrt{1 + \left(\dfrac{\partial h(x)}{\partial x}\right)^2}}.$$
The equation above in combination with Eq.~\ref{eq:arclength} correspond to the parametric equations for the local thickness $t(s)$. This, in turn, allows to find the local behaviour of the resistivity $\rho(t)$. 
%(using Eq.~\ref{eq:rho} below). 
The total charge resistance of the Al channel can be then calculated by using $R = \int_0^{L'} \rho(s)/[t(s) w] ds$.

%\subsection*{Model for spin transport in inhomogeneous curved channels}
	
A proper modelling of the charge and spin transport properties therefore requires to explicitly consider the thickness dependence of the resistivity \cite{steinhogl_size-dependent_2002}.  We do so by employing the Mayadas-Shatzkes (MS) model \cite{mayadas_electrical-resistivity_1970}, which accounts for the increase of electrical resistivity of the thin channel due to electron scattering at grain boundaries. Assuming that the thickness in the local surface normal of the Al channel corresponds to the smallest dimension between grain boundaries, the MS model provides us with a functional form of the resistivity as a function of the thickness, reading: 
\begin{equation} \label{eq:rho}
		\dfrac{\rho_0}{\rho(t)}= 3 \left[ \frac{1}{3} -\dfrac{\alpha}{2} + \alpha^2 - \alpha^3 \log{\left(1+ \frac{1}{\alpha}\right)} \right],  
\end{equation}
where $\rho_0$ is the resistivity of bulk Al, and $\alpha= \lambda_{\textbf{e}} \, C / [t (1-C)]$ can be determined from the knowledge of the electronic mean free path, $\lambda_{\textbf{e}}$, and the empirical reflectivity coefficient, $C$.  
We estimate the latter by using the value of the room-temperature mean free path $\lambda_{\textbf{e}} = 18.9$~nm and bulk Al resistivity $\rho_0 = 2.65  \times 10^{-8} \, \Omega\,$m \cite{gall_electron_2016}, and our experimental average resistivity at room temperature for reference Al channels of nominal thickness, $\rho(t_0) = 8.9 \times 10^{-8} \, \Omega\,$m.  We thereby obtain a reflectivity coefficient $C \simeq 0.82$. For the reference devices, we got a device to device statistical variance of $\approx 2~\Omega$, of the same size as the symbol for $A = 0$ in Fig.~\ref{fig:Experiment}(c). A statistical variance in the reflectivity coefficient of	$\pm 0.04$ allows us to account for this device to device variation.	Considering the scattering related to grain boundaries to be temperature independent, the obtained reflectivity coefficient can be further used to model the thickness dependent resistivity at low temperature, which we calibrate using our experimental average resistivity for reference channels at 4.2~K, $\rho(t_0) = 5.6 \times 10^{-8} \, \Omega\,$m. The values of resistivities considered above are consistent with the range of values observed for thin Al films in previous studies \cite{das_anisotropic_2016}. 
The ensuing behaviour of the charge resistance as a function of the trench height fits nicely 
with our experimental results [c.f. Fig.~\ref{fig:Experiment}(c)].

To obtain the inhomogeneous profile of the spin relaxation length, we use the fact that the latter can be expressed as
	%We recall that the spin relaxation length is 
	$\lambda=\sqrt{\tau_{\text{s}} D}$, where $D$ is the diffusion coefficient and $\tau_{\text{s}}$ is the spin relaxation time. Using the Einstein relation, $D=1/(\rho \, e^2\,N_{\text{Al}})$, with $N_{{\text{Al}}}$ the density of states in the channel at the Fermi level, we can therefore predict the thickness dependence of the diffusion constant. 
	Moreover, the Elliot-Yafet mechanism predicts
	%as mentioned above, spin relaxation in metallic Al films is dominated by the Elliott-Yafet mechanism \cite{kimura_temperature_2008,zutic_spintronics:_2004,jedema_spin_2003}, and thus 
	%we can predict 
	a scaling of the spin relaxation time, $\tau_{\text{s}} \propto \tau_{\text{p}} \propto 1/\rho$, where $\tau_{\text{p}}$ is the momentum relaxation time. 
	These considerations yield $\lambda \propto \rho^{-1}$, and allow to consider the ansatz for the thickness dependence of the spin relaxation length $\lambda(t)= \lambda_0 \rho_0 / \rho(t)$, 
	%\begin{equation}
	%	\lambda(t)= \lambda_0 \rho_0 / \rho(t), 
	%\label{eq:lambda}
	%\end{equation}
	whose functional form is uniquely determined by Eq.~\ref{eq:rho},
	%an ansatz for the thickness dependence of the spin relaxation length,
	%$$\lambda(t) = 3 \lambda_0 \left[ \frac{1}{3} -\dfrac{\alpha}{2} + \alpha^2 - \alpha^3 \log{\left(1+ \frac{1}{\alpha}\right)} \right],$$
	%where the functional form of $\alpha$ is uniquely determined by the charge transport model discussed above, 
	while the unknown $\lambda_0$ is fixed by requiring the spin relaxation length at the nominal thickness to be equal to that measured in reference devices, $\lambda_0=660$~nm at 4.2~K \cite{das_anisotropic_2016}. The ensuing spin relaxation length along the curved Al channel is shown in Fig.~\ref{fig:contour}(b), with a behaviour that is clearly inverse to that of the resistivity.

	%\textcolor{red}{The scaling of spin relaxation length mentioned above also allows to directly monitor the local behaviour of the spin current in our curved channel, which can be found to decay exponentially from the injector electrode with a functional form $\exp{\left[- \int^s 1/ \lambda(s^{\prime}) d s^{\prime}\right]} / \left[e \rho(s) \lambda(s)\right]$.
	%This also implies that the spin valve signal is determined by all the parameters discussed above, and can be expressed within the framework originally introduced by Takahashi and Maekawa \cite{takahashi_spin_2003}, but fully taking into account the inhomogeneity of the spin relaxation length along the channel.
	%}
	The spin valve signal is determined by the spin relaxation length and resistivity of the channel, which are both intrinsically inhomogeneous. This intrinsic inhomogeneity impedes the calculation of the spin signal using the simple analytical framework originally introduced by Takahashi and Maekawa for homogeneous channels \cite{takahashi_spin_2003}. 
	%On the other hand, 
	For this reason, 
	we have thereby extended the model by fully taking into account the inhomogeneity of the spin relaxation length along the channel [see Supporting Information Sections 2 and 3]. 
	With this approach, we find a closed expression for the spin accumulation signal in the ohmic contact regime, which reads:
	\begin{equation}
		\Delta R_{\textrm{NL}} = \dfrac{4 p_\textrm{F}^2}{(1-p_\textrm{F}^2)^2} \dfrac{{\cal R}_\textrm{F}^2}{{\cal R}_\textrm{N}} \dfrac{e^{-\int_0^{L'} \frac{1}{\lambda_\text{N}(s^{\prime})} ds^{\prime}}}{1 - e^{- 2 \int_0^{L'} \frac{1}{\lambda_\text{N}(s^{\prime})} ds^{\prime}}},
	\label{eq:analytical}
	\end{equation}
	where 
	%the spin relaxation length along the channel, $\lambda(s)$, is given by Eqs.~\ref{eq:rho} and \ref{eq:lambda}. 
	%Moreover, 
$w$ is the channel width, $L^{\prime}$ is the distance between injector and detector along the arclength \textit{\^{s}}, 
$\lambda_\textrm{N}$ is the equal spin relaxation length at the injector and detector, ${\cal R}_\textrm{N} = \rho_\textrm{N} \lambda_\textrm{N} / w t$,  ${\cal R}_\textrm{F}$ is the resistance of the ferromagnetic electrode with length $\lambda_\textrm{F}$ ($\lambda_\textrm{F}$ being the corresponding spin relaxation length), and $p_\textrm{F}$ is the current polarization of the ferromagnetic electrodes. The latter two quantities can be obtained from the spin signal in reference flat devices. Therefore, the knowledge of the local behaviour of the spin relaxation length allows us to obtain $\Delta R_{\textrm{NL}}$ as a function of the trench height. 
For the case of a homogeneous channel the integral in the exponents simplify to $L/\lambda_\textrm{N}$ and Eq.~\ref{eq:analytical} reproduces the usual theory \cite{takahashi_spin_2003}. 
By considering the same statistical variance in the reflectivity coefficient, $C$, derived from the charge transport above, we find a striking agreement between the theoretical results and the experimental spin valve data, as shown in Fig.~\ref{fig:Experiment}(b). The latter serves as experimental validation of our generalized diffusive spin transport model for inhomogeneous channels here presented. This in turn allows us to identify the dominant physical properties controlling spin transport in three-dimensional architectures, where inhomogeneity is directly controlled by the local geometry.

%\section*{Discussion}
%\section*{Independent geometrical control of spin and charge resistances}
%In particular, 
The analytical expression obtained in Eq.~\ref{eq:analytical} 
%leads to a compact approach to calculate spin signals for inhomogeneous channels where the Elliott-Yafet mechanism is dominant. 
%This 
allows us to interrogate in an efficient manner a broader phase space of geometrical variations of curved templates, e.g.\ as the Gaussian bumps described in Fig.~\ref{fig:contour}(a-b). The resulting 2D maps for charge resistance and spin resistance, due to the exploration of the phase space of Gaussian bump height $A$ and full width at half maximum $2\sqrt{2\log{2}}\sigma$, are shown in Fig.~\ref{fig:contour}(c-d). 
A key observation is the distinct scaling of the charge resistance and the spin resistance due to geometric control, evidenced by the different contour lines in both 2D maps. We highlight this difference by mapping a contour line from each 2D map into the other, resulting in 3D plots shown in Fig.~\ref{fig:contour}(e-f). Here, we observe the direct tuning of spin resistance independent of the charge resistance, and {\it vice versa}, via the nanoscale design of the template geometry. This hitherto unexplored approach to control the ratio of spin resistance to charge resistance, even within a single material system, has the potential to aid in the design of future circuits based on pure spin currents \cite{zutic_spintronics:_2004}.

Our curved-template approach enables control of the ratio of spin resistance to charge resistance in individual nanochannels, while allowing the fabrication of a spintronic architecture via a single deposition of the channel material. For flat homogeneous nanochannels the need of multiple deposition steps for each desired thickness would rapidly lead to an impractical fabrication process. Therefore, it is relevant to consider how simply tuning the length in flat homogeneous nanochannels, which is practical via lithography, compares with curved inhomogeneous nanochannels at the same nominal thickness. For a flat nanochannel to achieve a charge resistance $R/R^\text{ref} = 3.0$, its length must be increased to 3 times that of a reference channel, which leads to a spin resistance \cite{takahashi_spin_2003} of only $\Delta R_\text{NL}/\Delta R_\text{NL}^\text{ref} = 0.17$. This is significantly lower than the value of up to 0.52 obtained in Fig.~\ref{fig:contour}(f), and is one example of the general advantage offered by curved inhomogeneous channels for efficient individual control of spin and charge resistances [see Supporting Information Sections 4 and 5]. Spatial inhomogeneity below the characteristic length scale for spin transport, FWHM~$\lesssim \lambda$, combined with control of thickness down to the characteristic length scale for charge transport, $t \lesssim \lambda_e$, has been a hitherto unrecognised physical approach to enable such an efficient control within the context of Elliot-Yafet spin relaxation \cite{kimura_temperature_2008,zutic_spintronics:_2004,jedema_spin_2003}.

%\section*{Conclusions}

Using lateral non-local spin valves, we have demonstrated that an appropriate geometric design of metallic nanochannels yields spin resistance changes at constant electrical resistance and {\it vice versa}. Although spatially inhomogeneous nanochannels can be created in planar structures \cite{masourakis_modulation_2016}, our approach, using three-dimensional nanoarchitectures with a designed curved profile, intrinsically provides the necessary control to achieve the independent tuning of spin and charge resistances. Note that for planar structures there are other methods for controlling spin and charge currents. These rely on novel nanoscale materials, or heterostructures thereof, to gain functionality by active use of electric field \cite{guimaraes_controlling_2014, avsar_electronic_2016, avsar_gate-tunable_2017, lin_gate-driven_2017}, drift current \cite{ingla-aynes_eighty-eight_2016, vera-marun_nonlinear_2011} or proximity-induced spin relaxation \cite{yan_two-dimensional_2016, benitez_strongly_2018}. Such methods are highly relevant for current research, though their integration with current technologies is limited by their requirement of novel materials or low temperatures. 
%In contrast, our generally applicable geometrical approach does not depend on a specific material or heterostructure. 
In contrast, we expect our geometrical approach to be completely generic and thus applicable to other non-magnetic materials exhibiting a dominant Elliot-Yafet spin relaxation mechanism, e.g.\ Cu, or heterostructures thereof \cite{jedema_spin_2003, kimura_temperature_2008, masourakis_modulation_2016}. 
The combination of geometrical control and novel nanoscale materials is an interesting avenue for future spintronic technologies.

Recent works have explored technologically relevant curvilinear nanoarchitectures that transport vertically domain walls for magnetologic applications \cite{parkin_magnetic_2008,parkin_memory_2015}. Others have used curved templates pre-structured via self-assembly of nanostructures, which allows the nanoscale tuning of microstructure, thickness, and magnetic anisotropy of the deposited magnetic curved films \cite{streubel_magnetism_2016}. 
Geometrical effects can trigger new functionalities both in semiconducting \cite{gentile_edge_2015, nitta_2013, ying_designing_2016, gazibegovic_2017, chang_theoretical_2017} and superconducting \cite{turner_2010} low-dimensional systems. The geometrical control of pure spin currents demonstrated in this work can thus inaugurate the search for novel effects in spintronic devices using  other ultrathin curved materials like semiconductors and superconductors.

\section*{Associated content}

\noindent
\textbf{Supporting Information}\\
The Supporting Information is available free of charge on the ACS Publications website at http://pubs.acs.org.

\section*{Author information}

\noindent
\textbf{Corresponding Authors}\\
Correspondence and requests for materials should be addressed to K.S.D., C.O.\ and I.J.V.M.

\noindent
\textbf{Author Contributions}\\
C.O., P.G., D.M., I.J.V.M. conceived and supervised the project.
I.J.V.M.\ and K.S.D.\ conceived and designed the experiments. 
K.S.D.\ carried out the experiments and the initial data analysis.
D.M. supplied the curved templates and the corresponding electron microscopy imaging.
C.O.\ developed and performed the theoretical analysis with the help of P.G.\ and M.C.. K.S.D., C.O.\ and I.J.V.M.\ performed the final data analysis and wrote the manuscript with the help of B.J.v.W., M.C., P.G., and D.M. All authors discussed the results and the manuscript.

\noindent
\textbf{Notes}\\
The authors declare no competing financial interest.

\section*{Acknowledgments}

\noindent
We thank Dr.\ S. Baunack (IFW Dresden) for his contribution at the initial stage of this work and J.\ G.\ Holstein, H.\ M.\ de Roosz, H.\ Adema and T.\ Schouten for their technical assistance. Support by the Structural Characterization Facilities at IBC of the HZDR is gratefully acknowledged. We acknowledge the financial support of the Zernike Institute for Advanced Materials and the Future and Emerging Technologies (FET) programme within the Seventh Framework Programme for Research of the European Commission, under FET-Open Grant No.~618083 (CNTQC). C.O.\ acknowledges support from the Deutsche Forschungsgemeinschaft (Grant No.\ OR 404/1-1), and from a VIDI grant (Project 680-47-543) financed by the Netherlands Organization for Scientific Research (NWO). D.M. acknowledges support from the ERC within the EU 7th Framework Programme (ERC Grant No. 306277).

\section*{References}
%\bibliographystyle{achemso}
%\bibliography{curvaturepaper}

\providecommand{\latin}[1]{#1}
\makeatletter
\providecommand{\doi}
  {\begingroup\let\do\@makeother\dospecials
  \catcode`\{=1 \catcode`\}=2\doi@aux}
\providecommand{\doi@aux}[1]{\endgroup\texttt{#1}}
\makeatother
\providecommand*\mcitethebibliography{\thebibliography}
\csname @ifundefined\endcsname{endmcitethebibliography}
  {\let\endmcitethebibliography\endthebibliography}{}

\clearpage

%\appendix

%\renewcommand{\figurename}{Supplementary FIG.}
\renewcommand{\thesection}{\arabic{section}}
\renewcommand{\theequation}{S\arabic{equation}}%   
\newcommand{\Z}{\mathbb{Z}}

\renewcommand{\thefigure}{S\arabic{figure}}%   
\setcounter {equation} {0}
\setcounter {figure} {0}

\section*{Supporting Information}

\section{Room temperature measurements}

\begin{figure}[bp]
	\includegraphics[width=0.95\linewidth]{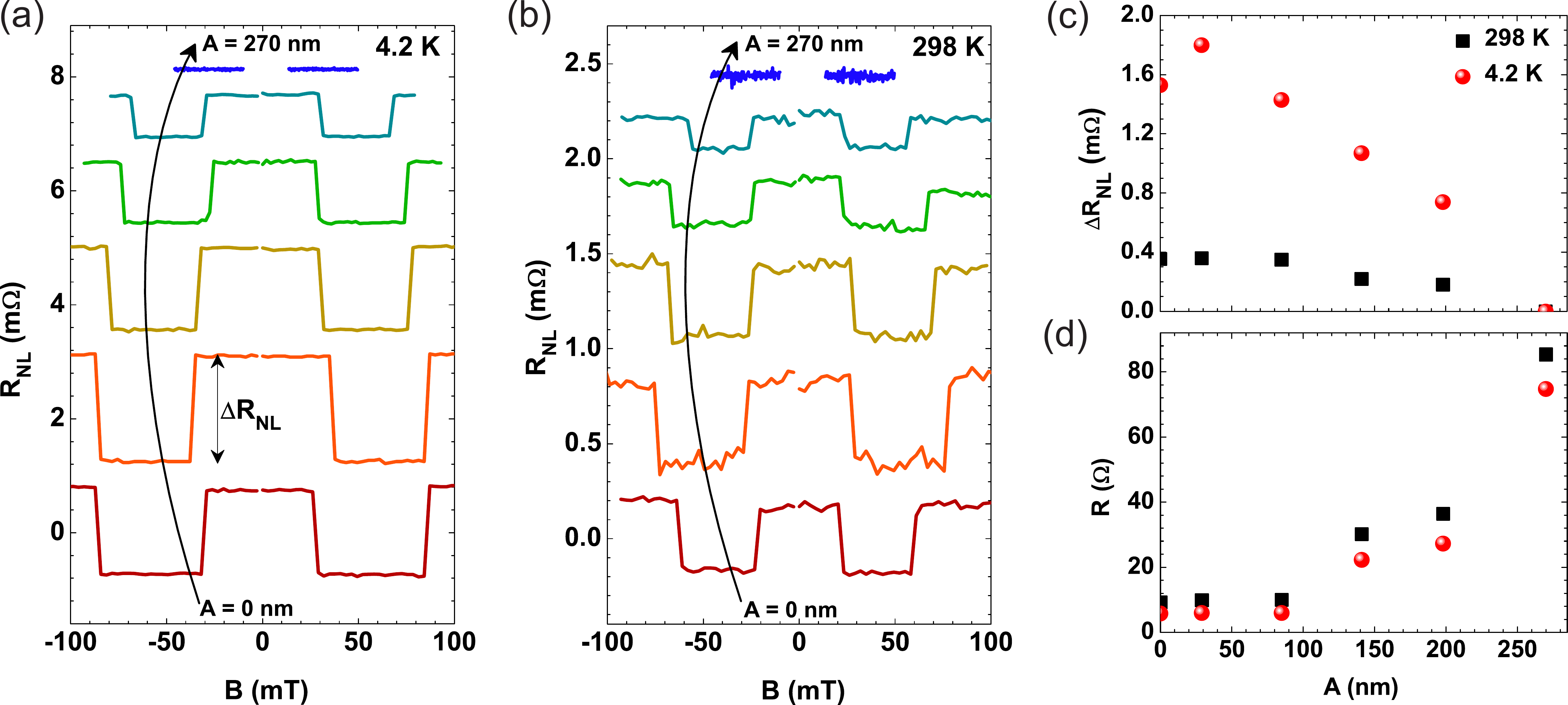}
	\caption{
		\label{fig:RT}
		\textbf{Measurements at both room temperature and at 4.2~K.}
		\textbf{(a-b)} Non-local spin valve measurements at (a) $T=4.2$~K, and (b) $T=298$~K for spin valves with different curvatures of the Al channel, corresponding to different trench heights ($A$). The black arrows represent the increasing direction of $A$. The spin valve signal $\Delta R_{\text{NL}}$ decreases with increasing $A$. \textbf{(c)} $\Delta R_{\text{NL}}$ as a function of $A$, with $A=0$ representing the reference flat devices. \textbf{(d)} The charge resistance ($R$) of the Al channel between the injector and detector electrodes plotted against $A$.
	}
\end{figure}

	Besides the spin valve measurements at 4.2~K shown in the main text, we have also performed measurements at room temperature (298~K). A direct comparison between both sets of measurements is shown in Fig.~\ref{fig:RT}(a-b). The behaviour at both temperatures is consistent, with $\Delta R_{\text{NL}}$ being maximum for the reference spin valves with $A=0$ and showing a significant decrease for trench heights above $\approx 100$~nm. Although the variation with trench height is similar for both temperatures, it is visible that the spin valve signal at room temperature is lower than at 4.2~K, see Fig.~\ref{fig:RT}(c). We note that this change is driven by the decrease in spin relaxation length with increasing temperature. The latter was measured for the reference 50~nm thick flat devices to be $\lambda = 380$~nm at room temperature, whereas it was 660~nm at 4.2~K, as mentioned in the main text \cite{das_anisotropic_2016}.

	A similar response at both temperatures was also observed for the four-probe charge resistance ($R$) of the curved channel between the injector and detector electrodes, showing a steep increase in resistance for trenches with height greater than $\approx 100$~nm, as observed in Fig.~\ref{fig:RT}(d). We note that our theoretical modelling captures well the measured charge resistances of our samples at both temperatures, using a unique reflectivity coefficient $C=0.82 \pm 0.04$, which allows to correctly predict the charge resistances of the reference flat devices. We emphasize that the fact that the charge resistances of the flat devices at both room temperature and low temperature are within this reflectivity coefficient variance provides additional evidence of the temperature independence of the reflectivity coefficient. 
	
\begin{figure}[htbp]
	\includegraphics[width=0.9\linewidth]{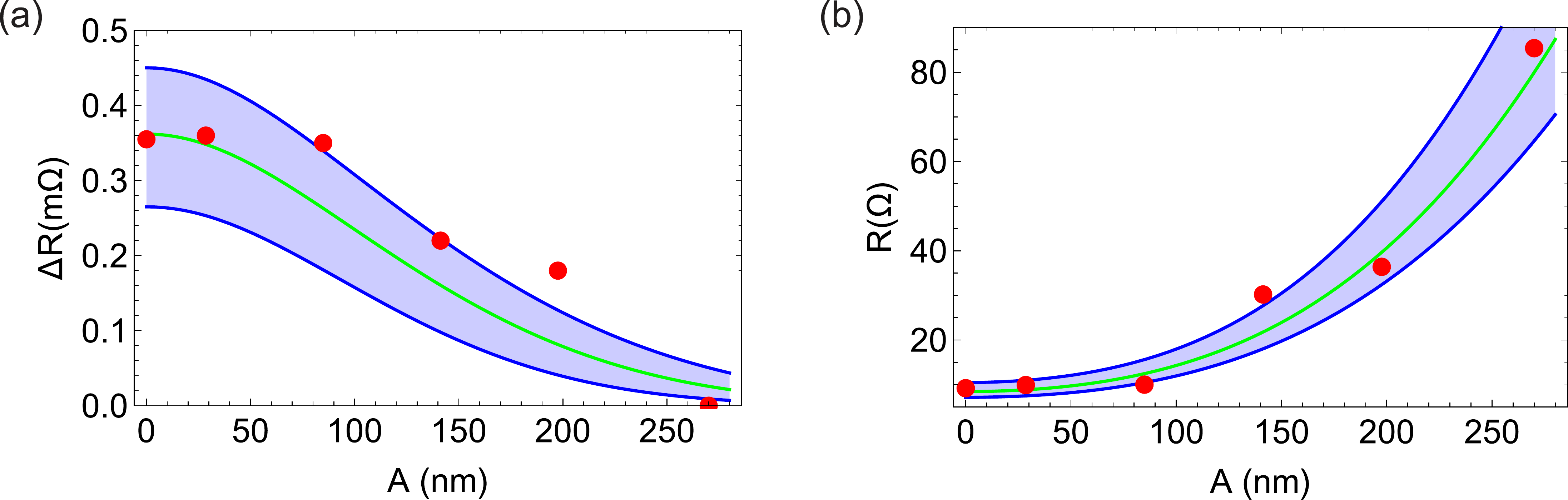}
	\caption{
		\label{fig:RTcomparison}
		\textbf{Room temperature measurements and modelling.}
		The spin signal \textbf{(a)} and the charge resistance \textbf{(b)} as a function of the trench height $A$ at room temperature. The experimental data and the theory results are shown as points and lines, respectively. The shaded region represents the uncertainty due to device-to-device variation.}
\end{figure}

The agreement between our calculations and the experimental spin resistances for both flat and trench devices, both at low temperature and room temperature, validates our modelling of the Elliot-Yafet mechanism within the condition $\lambda \propto 1/\rho$. In particular, the relation is better modelled to couple spin and momentum relaxation rates for discrete scattering mechanisms, with possibly separate scaling factors for phonon and temperature-independent scatterers \cite{villamor_contribution_2013}. In practice, these scaling factors are observed to be similar. More importantly, at low temperature our model is certainly valid as phonon scattering is suppressed. The observation of a good agreement also at room temperature is a consequence of the ultrathin nature of our nanochannels, where grain (surface) boundary scattering is a dominant mechanism. The comparison between the theory analysis and the experimental results for both the charge and spin resistance at room temperature is shown in Fig.~\ref{fig:RTcomparison}.

\section{Pure spin currents in inhomogeneous metallic channels}
The description of spin and charge transport in a thin metallic channel with a spatially varying electrical conductivity can be derived starting out from the continuity equations for charge and spin in the steady state:
\begin{eqnarray*}
\nabla \cdot \left({\bf j}_{\uparrow} + {\bf j}_{\downarrow}\right) &=&0 \\
\nabla \cdot \left({\bf j}_{\uparrow} - {\bf j}_{\downarrow}\right) &=& -e \dfrac{\delta n_{\uparrow}}{\tau_{\uparrow \downarrow}} + e \dfrac{\delta n_{\downarrow}}{\tau_{\downarrow \uparrow}},
\end{eqnarray*}
where $\tau_{\sigma \sigma^{\prime}}$ is the scattering time of an electron from spin state $\sigma$ to $\sigma^{\prime}$, and $\delta n_{\sigma}$ is the carrier density deviation from equilibrium in the $\sigma$ spin channel. The electrical current in each spin channel can be related, as usual, to the gradient of the electrochemical potential via ${\bf j}_{\sigma} = -(\sigma_{\sigma}/e) \nabla \mu_{\sigma}$, with $\sigma_{\sigma}$  the local electrical conductivity. 
Using the detailed balancing relation between the scattering times and the density of states in each spin sub-band $N_{\uparrow} / \tau_{\uparrow \downarrow} = N_{\downarrow} / \tau_{\downarrow \uparrow}$, we obtain the following two equations for the electrochemical potentials:
\begin{eqnarray} 
& &\nabla \cdot \left[ \sigma_{\uparrow}(s) \nabla \mu_{\uparrow}(s) +  \sigma_{\downarrow}(s) \nabla \mu_{\downarrow}(s)  \right]  \equiv   0 \label{eq:steadycurrent}   \\
& &\nabla^{2} \left(\mu_{\uparrow}(s) - \mu_{\downarrow}(s) \right) + \dfrac{\nabla \sigma_{\uparrow}(s)}{\sigma_{\uparrow}(s)} \cdot \nabla \mu_{\uparrow}(s) - \dfrac{\nabla \sigma_{\downarrow}(s)}{\sigma_{\downarrow}(s)} \cdot \nabla \mu_{\downarrow}(s)  \equiv  \nonumber \\ & & \dfrac{1}{\lambda^{2}(s)} \left(\mu_{\uparrow}(s) - \mu_{\downarrow}(s) \right) 
  \label{eq:steadyspin}
\end{eqnarray}	
In the equations above we introduced the spin relaxation length $\lambda$, which, as the electrical conductivity, depends on the channel coordinate $s$. 
Eqs.~\ref{eq:steadycurrent},\ref{eq:steadyspin} generalize the equations for an homogeneous channel reported in Ref.~\onlinecite{takahashi_spin_2003}, and can be generally solved by resorting to numerical methods. However, for a nonmagnetic channel where $\sigma_{\downarrow}(s) \equiv \sigma_{\uparrow}(s)$ and considering a pure spin current ${\bf j}_{\downarrow}(s) + {\bf j}_{\uparrow}(s) \equiv 0$ , it is possible to analytically express the local behaviour of the electrochemical potentials assuming that spin relaxation is dominated by the Elliot-Yafet mechanism. Since in this situation $\lambda(s) \propto \sigma(s)$ [see the main text], Eq.~\ref{eq:steadyspin} is transformed as 
\begin{eqnarray*}
& & \nabla^2 \left(\mu_{\uparrow}(s) - \mu_{\downarrow}(s) \right) + \dfrac{\nabla \lambda(s)}{\lambda(s)} \cdot \nabla  \left(\mu_{\uparrow}(s) - \mu_{\downarrow}(s) \right) \equiv \\ & & \dfrac{1}{\lambda^{2}(s)} \left(\mu_{\uparrow}(s) - \mu_{\downarrow}(s) \right), 
\end{eqnarray*}
which is satisfied for   
$$ \left(\mu_{\uparrow}(s) - \mu_{\downarrow}(s) \right)  \equiv \mathrm{e}^{\displaystyle \pm \int^{s} \dfrac{1}{\lambda(s^{\prime})} d s^{\prime}}.$$ 
As detailed in the next section, this knowledge of the local behaviour of the electrochemical potential allows to express the spin accumulation signal in a non-local spin valve as in Eq.~2 of the main text.

\section{Spin accumulation signal}
In order to derive the expression for the spin-dependent voltage in our inhomogeneous non-local spin valves, we start out by considering that when the bias current $I$ flows from the ferromagnetic injector at $s \equiv 0$ to the ``left" side of the normal channel ($s<0$), the solution for the electrochemical potentials in the inhomogeneous normal channel can be written as 
\begin{eqnarray*} 
\mu_{\beta}^{\text{N}}(s) &=& \dfrac{e I}{\sigma^{\text{N}}_0 A_\text{N}} s + \beta \left[ a_1 \,\mathrm{e}^{\dfrac{s}{\lambda_0}} + a_2 \, \mathrm{e}^{\dfrac{s}{\lambda_0} - \displaystyle \int_{0}^{L} \dfrac{1}{\lambda(s^{\prime})} d s^{\prime}}  \right] \hspace{.5 cm} s \leq 0 \\
\mu_{\beta}^{\text{N}}(s) &=& \beta \left[ a_1 \,\mathrm{e}^{- \displaystyle \int_{0}^{s} \dfrac{1}{\lambda(s^{\prime})} d s^{\prime}} + a_2 \, \mathrm{e}^{\displaystyle \int_{0}^{s} \dfrac{1}{\lambda(s^{\prime})} d s^{\prime} - \displaystyle \int_{0}^{L} \dfrac{1}{\lambda(s^{\prime})} d s^{\prime}}  \right] \hspace{.5 cm} 0 \leq s \leq L \\
\mu_{\beta}^{\text{N}}(s) &=& \beta \left[ a_1 \,\mathrm{e}^{- \dfrac{s-L}{\lambda_0} - \displaystyle \int_{0}^{L} \dfrac{1}{\lambda(s^{\prime})} d s^{\prime}} + a_2 \, \mathrm{e}^{-\dfrac{s - L}{\lambda_0}}  \right] \hspace{.5 cm}  s \geq L 
\end{eqnarray*}  
In the equations above, $\lambda_0$ and $\sigma_0^{\text{N}}$ are respectively the constant spin relaxation length and the electrical conductivity in the two homogeneous regions exterior to the ferromagnetic injector and detector, which, for simplicity, are taken to be equal in magnitudes. Moreover,  $I$ is the charge current flowing from the injector to the left end of the normal channel, and $L$ is the actual distance among the two ferromagnets. Finally the index $\beta= \pm 1$ indicates the two spin channels, and $A_\text{N}$ is the channel cross-sectional area. 

In the ferromagnetic electrodes, the thickness is much larger than the spin relaxation length, and thus the solutions close to the interface take the forms of vertical transport along the $z$ direction: $\mu^{\text{F1,F2}}_{\beta}= \overline{\mu}^{\text{F1,F2}} + \beta (\sigma^{\text{F}} b_{1,2} / \sigma^{\text{F}}_{\beta}) \exp{(-z / \lambda_\text{F})} $. Here, $\overline{\mu}^{\text{F1}}= (e I / \sigma^\text{F} A_\text{F}) z + e V_{1}$  describes the charge current flow in the ferromagnetic injector, with $A_\text{F}$ the corresponding cross section and $\sigma^\text{F}= \sigma^\text{F}_{\uparrow} + \sigma^\text{F}_{\downarrow}$ 
the total electrical conductivity, while $\overline{\mu}^{\text{F2}}= e V_{2}$ is the constant potential drop which changes sign when the injector and detector magnetizations change from parallel to antiparallel. 

To proceed further, we treat the interfacial currents across the junctions assuming transparent metallic contacts, and thereby require the continuity of the spin-dependent electrochemical potentials. With this assumption, the six equations for the two potential drops and the four parameters $a_{1,2}, b_{1,2}$ explicitly read: 
\begin{eqnarray*}
p_\text{F} I &\equiv & \dfrac{2 \,a_{1}}{e \mathcal{R}_{\text{N}}} + \dfrac{2 \,b_{1}}{e \mathcal{R}_{\text{F}}} \\ \\
e V_{1} + \dfrac{2 \beta b_1}{(1 + \beta p_\text{F})} & \equiv & \beta \left[a_1 + a_2  \mathrm{e}^{- \displaystyle \int_{0}^{L} \dfrac{1}{\lambda(s^{\prime})} d s^{\prime}} \right] \\ \\ 
0 &\equiv & \dfrac{2 \,a_{2}}{e \mathcal{R}_{\text{N}}} + \dfrac{2 \,b_{2}}{e \mathcal{R}_{\text{F}}} \\ \\
e V_{2} + \dfrac{2 \beta b_2}{(1 + \beta p_\text{F})} & \equiv & \beta \left[a_1 \mathrm{e}^{- \displaystyle \int_{0}^{L} \dfrac{1}{\lambda(s^{\prime})} d s^{\prime}} + a_2 \right] 
\end{eqnarray*} 
In the equations above we have introduced the current polarization $p_\text{F}$ of the injector and detector electrodes, the resistance of the ferromagnetic electrodes over the spin relaxation length distance, and the resistance of the normal channel ${\mathcal R}_{\text{N}}= \lambda_0 / (\sigma_0^\text{N} A_\text{N})$, which is a constant in the Elliot-Yafet framework. 
Solving the equations for the spin-dependent voltage, and assuming ${\mathcal R}_{\text{F}} \ll {\mathcal R}_\text{N}$, we thereby obtain:
\begin{equation*}
\dfrac{V_{2}}{I}=\pm  \dfrac{2 p_\text{F}^2}{\left(1-p_\text{F}^2\right)^2} \dfrac{{\mathcal R}_\text{F}^2}{{\mathcal R}_\text{N}} \times \dfrac{\mathrm{e}^{- \displaystyle \int_0^L \dfrac{1}{\lambda(s^{\prime})} d s^{\prime}}}{1 - \mathrm{e}^{- 2 \displaystyle \int_0^L \dfrac{1}{\lambda(s^{\prime})} d s^{\prime}}}.
\end{equation*}

\section{Effect of changing the total thickness and/or the channel length of a flat homogeneous channel}

In the main part of the manuscript, we have shown the possibility to achieve an independent tuning of the spin and charge resistances utilizing the curved geometry of a nanochannel. This has been done by comparing the spin and charge resistances of an Al nanochannel deposited on a trench substrate with respect to the resistances of a conventional flat nanochannel with fixed total length and width. 
In this section we show that, although an independent tuning of spin and charge resistances can be also achieved in conventional geometries assuming the length of the nanochannel can be varied at wish, this approach is inefficient. 
As we will show below, the advantage of using curved nanochannels relies on the fact that the inhomogeneous behaviour of the resistance generally yields larger spin signals. 

To start with, let us consider a conventional flat nanochannel. In the remainder, and for simplicity, we will always assume that the width of the nanochannel is kept constant while the other structural parameters can be independently tuned. The spin accumulation signal for metallic contacts is generally given by the well-known formula of Takahashi and Maekawa \cite{takahashi_spin_2003} that reads: 
\begin{equation} 
	\Delta R_{\text{NL}} = \dfrac{4 p_\text{F}^2}{\left(1-p_\text{F}^2\right)^2} \, \dfrac{\mathcal R_\text{F}^2}{\mathcal R_\text{N}} \, \dfrac{\mathrm{e}^{-\frac{L}{\lambda}}}{1-\mathrm{e}^{-\frac{2 L}{\lambda}}},
\end{equation}
where, $p_\text{F}$ is the equal current polarizations of the ferromagnetic injector and the detector and $L$ is the distance between the injector and the detector. Moreover, $\mathcal R_\text{N}$ ($\mathcal R_\text{F}$) is the resistance of the Al channel (ferromagnetic injector and the detector) with a cross-sectional area $A_\text{N}$ ($A_\text{F}$) and length equal to one spin relaxation length $\lambda$ ($\lambda_\text{F}$). Therefore, $\mathcal R_\text{N}=\rho \lambda / (w t)$, where $w$, $t$ and $\rho$ are the width, thickness and resistivity of the nanochannel, respectively. When allowing for arbitrary changes in the thickness of the channel, the value of $\mathcal R_\text{N}$ changes not only via the thickness but also via the corresponding changes in the resistivity and consequently in the spin relaxation length. 

We now aim to compare how the scaling of charge and spin resistances evolve with geometry for the case of a flat homogeneous channel, where we only change the thickness of the whole channel and/or the length of the channel between the injector and the detector electrodes. We again use the Elliot-Yafet framework, $\lambda(t) \propto \rho(t)^{-1}$, that implies $\rho(t) \lambda(t) = \textrm{const}$. With this, we can define a reference resistance $\mathcal R_\text{N}^0=\rho_0 \lambda_0 / (w t_0)$, where $\rho_0$ and $\lambda_0$ are the resistivity and spin relaxation length at a reference thickness $t_0$, respectively. We next employ this in the expression for the spin signal and rewrite it as 
\begin{equation} 
	\dfrac{\Delta R_{\text{NL}}}{\mathcal R_\text{N}^0} = \dfrac{4 p_\text{F}^2}{\left(1-p_\text{F}^2\right)^2} \, \left(\dfrac{\mathcal R_\text{F}}{\mathcal R_\text{N}^0}\right)^2 \, \dfrac{t}{t_0} \dfrac{\mathrm{e}^{-\frac{L}{\lambda}}}{1-\mathrm{e}^{-\frac{2 L}{\lambda}}}.
\end{equation}
Henceforth, it follows that the equation above provides us with the functional form of the spin signal for a nanochannel with varying thickness and length. 
Since, as mentioned above, the spin relaxation length is also thickness dependent, we overcome this additional structural parameter dependence as follows. First we notice that the total charge resistance of a flat homogeneous channel of length $L$ is simply given by $R=\rho L / (t w)$, from which we can read off the ratio between the length $L$ and the spin relaxation length $\lambda$ as 
\begin{equation}
\dfrac{L}{\lambda}=\dfrac{R}{\mathcal R_\text{N}}= \dfrac{R}{\mathcal R_\text{N}^0} \dfrac{t}{t_0}.
\end{equation}

As a final result, we can express the spin signal in the form 
\begin{equation} 
	\dfrac{\Delta R_{\text{NL}}}{\mathcal R_\text{N}^0} = \mathcal{M}^2\, \dfrac{t}{t_0}\, \dfrac{\mathrm{e}^{-\frac{R}{\mathcal R_\text{N}^0} \, \frac{t}{t_0}}}{1- \mathrm{e}^{-2 \frac{R}{\mathcal R_\text{N}^0} \, \frac{t}{t_0}}},
	\label{FlatChannelEq} 
\end{equation}
where we have introduced the factor $\mathcal{M}^2=4 p_F^2 \left(\mathcal R_\text{F} / \mathcal R_\text{N}^0 \right)^2 / \left(1- p_F^2\right)^2$ that depends on the properties of the ferromagnetic injector and detector and the reference resistance $\mathcal R_\text{N}^0$. Eq.~\ref{FlatChannelEq} implies that for a fixed nanochannel charge resistance $R$, the spin signal can be modified by changing the thickness or alternatively the channel length. 

\begin{figure}[tbp]
	\begin{center}
		\includegraphics[width=0.9\linewidth]{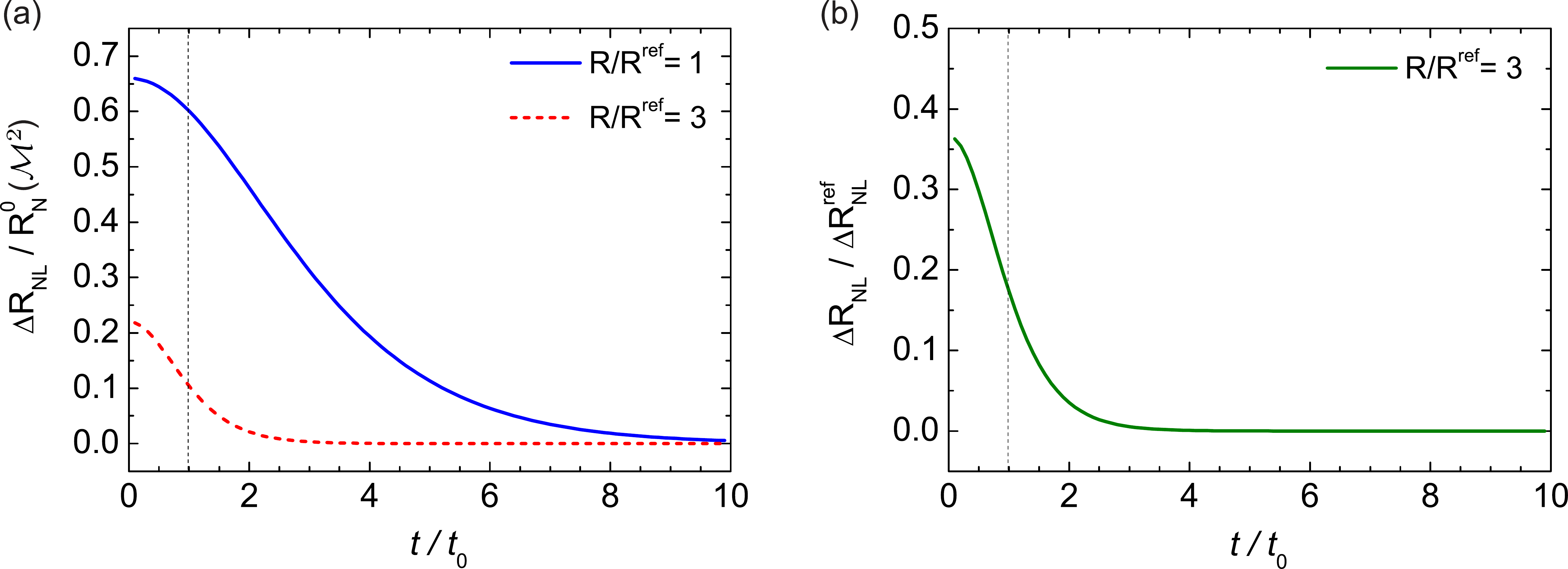}
	\end{center}
	\caption{\textbf{Spin signal for a flat homogeneous channel.} \textbf{(a)} Spin signal $\Delta R_{\text{NL}} / \mathcal R_\text{N}^0$ in units of the parameter $\mathcal{M}^2$ plotted as a function of $t/t_0$ for a flat homogeneous nanochannel for two different values of the charge resistance $R/R^\text{ref}=1$ (solid blue line) $R/R^\text{ref}=3$ (dashed red line). \textbf{(b)} Modulation of the spin signal in a flat homogeneous channel as a function of $t/t_0$, when the total charge resistance of the channel is 3 times of that of a reference flat channel.}
	\label{fig:flatchannel}
\end{figure}

We therefore consider a reference channel length, thickness and spin relaxation length of 500~nm, 50~nm and 660~nm, respectively (exactly the same parameters as the reference devices in the main text). Next, we plot the corresponding behaviour of the spin signal in Fig.~\ref{fig:flatchannel}(a), using Eq.~\ref{FlatChannelEq}, for two distinct cases. First, corresponding to the case when the total charge resistance of the flat channel is equal to the reference charge resistance, i.e.\ $R=R^\text{ref}=0.76\mathcal R_\text{N}^0$. Second, for the case when the total charge resistance of the channel is now made 3 times that of the reference resistance, i.e.\ $R=3R^\text{ref}=2.27\mathcal R_\text{N}^0$, where it is apparent the spin signal is suppressed for an equal thickness.

A direct comparison of the spin signal in flat homogeneous channels for the two cases considered above is shown in Fig.~\ref{fig:flatchannel}(b). 
Here we observe that, although full tuning of both thickness and length in flat homogeneous channels can lead to control of spin resistance (at a fixed charge resistance), the obtained spin resistance values are strictly lower than those from curved inhomogeneous nanochannels. This is evident by comparison with the result shown in Fig.~3(d) in the main text.

Furthermore, our curved-template approach enables controlling the ratio of spin resistance to charge resistance in individual nanochannels, while allowing the fabrication of a spintronic architecture via a single deposition step of the channel material. On the other hand, for an spintronic architecture based on flat homogeneous nanochannels, the need of multiple deposition steps for each desired thickness rapidly scales to a fabrication process impractical to implement.

Therefore, it is relevant to consider how tuning only the length in flat homogeneous nanochannels compares with curved inhomogeneous nanochannels, at the same nominal thickness. At $t=t_0$, to tune the charge resistance to $R=3R^\text{ref}$, the length of a flat nanochannel has to be increased to 3 times that of the reference channel. This results in a spin resistance of only 0.17 times that of a reference channel, as indicated by the vertical dotted line in Fig.~\ref{fig:flatchannel}(b). On the other hand, we find that for the same nominal thickness and $R=3R^\text{ref}$ condition, a curved inhomogeneous nanochannel leads to a spin signal of up to 0.52 times that of a reference channel (see Fig.~3(d) in the main text). This is a clear example of the advantage offered by curved inhomogeneous channels towards an efficient tuning the ratio of spin resistance to charge resistance.

\section{Generalized advantage of a curved inhomogeneous nanochannel}

We next investigate the corresponding change of the spin signal in a curved nanochannel where, as before, we assume to vary the total arclength and thickness of the non-magnetic material. We recall that the functional form of the spin signal derived in the main part of the manuscript for a curved inhomogeneous nanochannel reads 

\begin{equation}
\Delta R_{\textrm{NL}} = \dfrac{4 p_\textrm{F}^2}{(1-p_\textrm{F}^2)^2} \dfrac{{\cal R}_\textrm{F}^2}{{\cal R}_\textrm{N}} \dfrac{e^{-\int_0^{L'} \lambda_\text{N}^{-1}(s) ds}}{1 - e^{- 2 \int_0^{L'} \lambda_\text{N}^{-1}(s) ds}},
\label{eq:analytical}
\end{equation}
where, $\mathcal R_\text{N}= \rho_\text{N} \lambda_\text{N} / (w t)$, as explained in the main text. The thickness and the resistivity of the channel on the left of the injector and on the right of the detector are assumed to be constant and given by $t$ and $\rho_\text{N}$, respectively. Moreover, $\lambda_\text{N}$ is the spin relaxation length corresponding to the resistivity $\rho_\text{N}$. Using simple algebra, the equation above can be recast in the following form: 
\begin{equation} 
	\dfrac{\Delta R_{\text{NL}}}{\mathcal R_\text{N}^0} = \mathcal{M}^2\, \dfrac{t}{t_0}\,\dfrac{\mathrm{e}^{-\int_0^L \lambda_\text{N}^{-1}(s)ds}}{1-\mathrm{e}^{-2 \int_0^L \lambda_\text{N}^{-1}(s)ds}}.
\end{equation}

\begin{figure}[b]
	\begin{center}
		\includegraphics[width=0.92\linewidth]{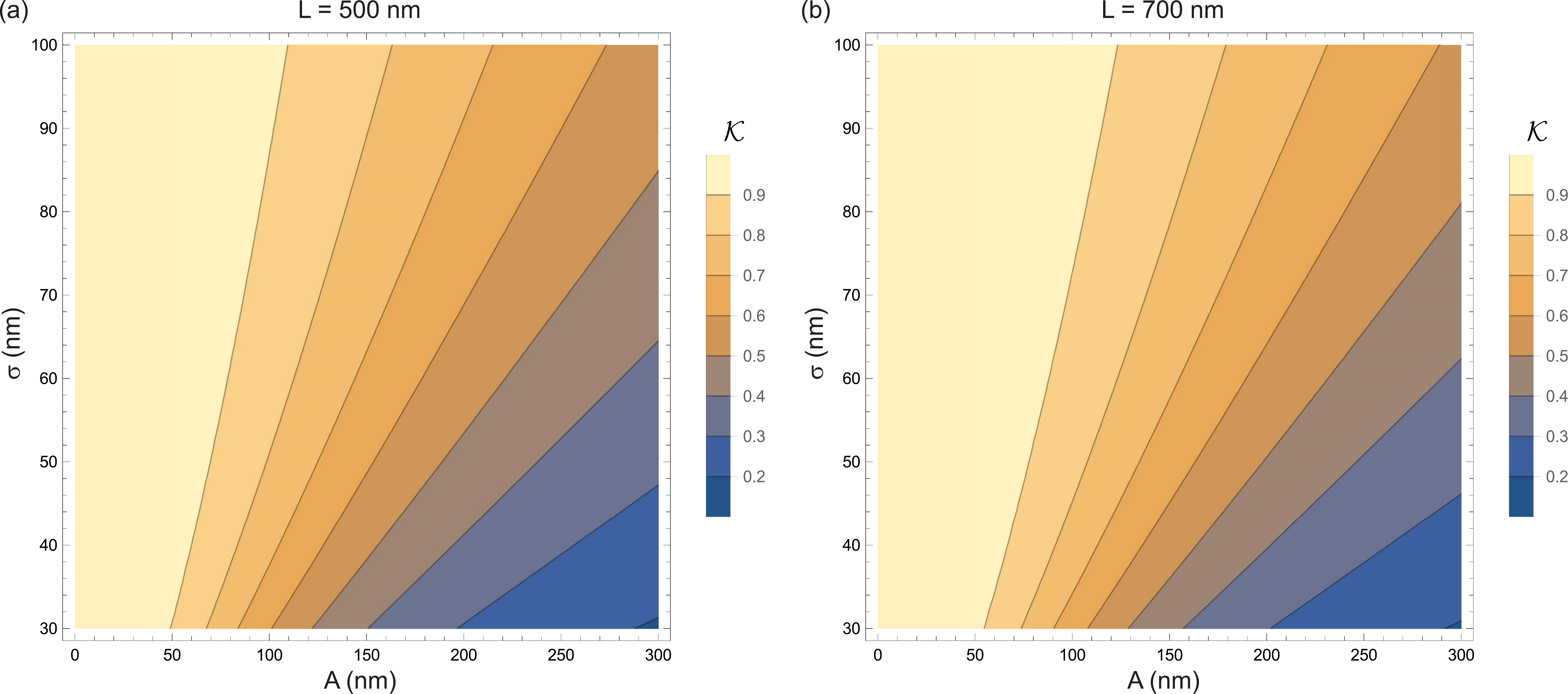}
	\end{center}
	\caption{\textbf{Dependence of the curved factor $\mathcal K$ on the curved channel geometry.} Behaviour of the $\mathcal{K}$ factor for a Gaussian bump with height $A$ and full width at half maximum $2\sqrt{2\log{2}}\sigma$, for injector-detector separation $L=500$~nm \textbf{(a)} and $L=700$~nm \textbf{(b)}.}
	\label{CurvedFactor}
\end{figure}

To obtain the behaviour of the spin signal for a fixed charge resistance, we need to introduce the total charge resistance of the curved channel, that simply reads 
\begin{equation} 
	R=\int_0^L \dfrac{\rho_\text{N}(s)}{w t(s)}ds.
\end{equation}
It is also straightforward to show that in the Elliot-Yafet mechanism the ratio between the total charge resistance and the characteristic reference spin resistance $\mathcal R_\text{N}^0$ can be expressed as 
\begin{equation}
	\dfrac{R}{\mathcal R_\text{N}^0}=\int_0^L \dfrac{1}{\lambda_\text{N}(s)} \, \dfrac{t_0}{t(s)}ds. 
\end{equation}
This equation can be used in order to write 
\begin{equation} 
	\int_0^L \dfrac{1}{\lambda_\text{N}(s)}ds = \dfrac{R}{\mathcal R_\text{N}^0} \dfrac{\int_0^L \lambda_\text{N}(s)^{-1}ds}{\int_0^L \lambda_\text{N}(s)^{-1} \times t_0 / t(s)ds},
\end{equation}
which can be simplified as 
\begin{equation}
	\int_0^L \dfrac{1}{\lambda_\text{N}(s)}ds= \mathcal{K} \dfrac{R}{\mathcal R_\text{N}^0} \, \dfrac{t}{t_0},
\end{equation}
where, we have introduced the curved factor 
\begin{equation}
	\mathcal{K} = \dfrac{\dfrac{1}{L} \mathlarger{ \int_0^L} \dfrac{\lambda_0}{\lambda_\text{N}(s)} \dfrac{t_0}{t}ds}{\dfrac{1}{L} \mathlarger{\int_0^L} \dfrac{\lambda_0}{\lambda_\text{N}(s)} \dfrac{t_0}{t(s)}ds}. 
\end{equation}
Clearly, for a conventional flat channel, the curved factor $\mathcal{K}$ reduces to one. Moreover, we can express the generic form of the spin signal simply as 
\begin{equation} 
	\dfrac{\Delta R_{\text{NL}}}{\mathcal R_\text{N}^0} = \mathcal{M}^2\, \dfrac{t}{t_0}\, \dfrac{\mathrm{e}^{- \mathcal{K} \, \frac{R}{\mathcal R_\text{N}^0} \, \frac{t}{t_0}}}{1- \mathrm{e}^{-2 \,\mathcal{K} \, \frac{R}{\mathcal R_\text{N}^0} \, \frac{t}{t_0}}}.
	\label{spin_signal_compare} 
\end{equation}
Eq.~\ref{spin_signal_compare} allows us to directly compare the spin signal of a flat channel with a given charge resistance and thickness to that of an inhomogeneous channel with the same total charge resistance. Clearly, whenever the curved factor $\mathcal{K}<1$, there is a gain in the spin signal even though the charge resistance is the same. Fig.~\ref{CurvedFactor} shows that this is indeed the case. Therefore, the advantage of using the inhomogeneity of a curved channel is a generic gain in the spin signal with respect to the flat channel case.

\section{Simplified model only considering an increased channel length}

The data in Fig.~2 in the main text was described with our extended model of spin and charge transport in a thin metallic channel with a spatially varying electrical conductivity. It is illustrative to compare the same data with a simplified model that only considers the increase in the channel length due to the curved geometry, ignoring any variation in the thickness of the Al channel, in order to see the difference with respect to the extended model. 

\begin{figure}[h]
	\includegraphics[width=0.9\linewidth]{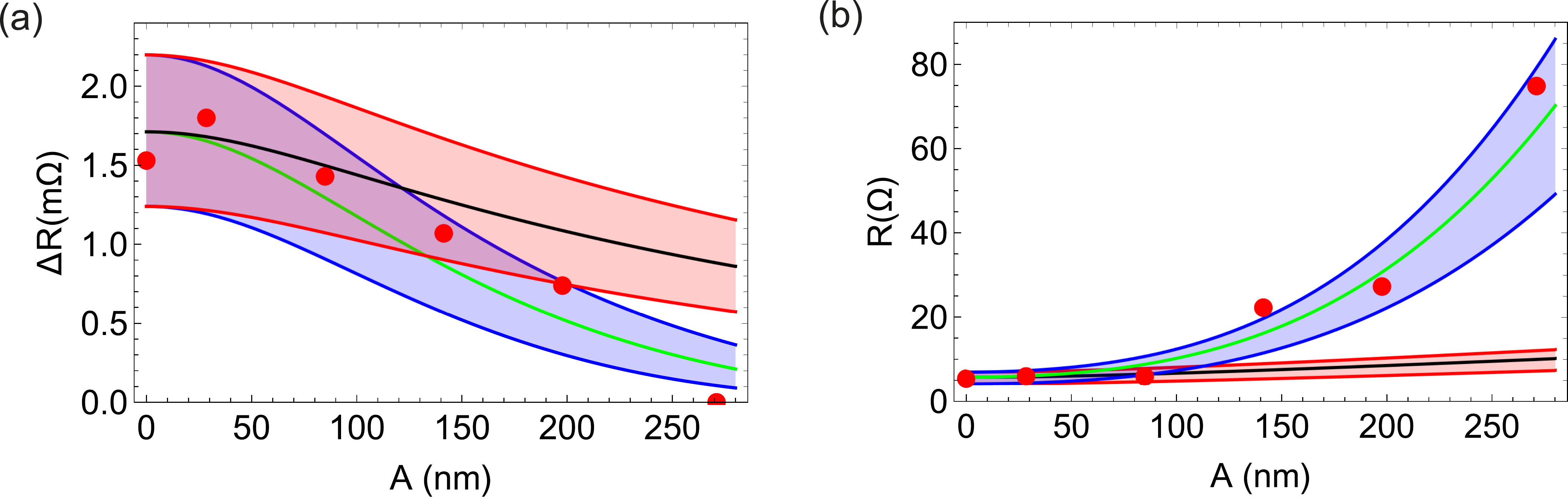}
	\caption{
		\label{fig:LTsimple}
		\textbf{Comparison with simplified model.}
		The spin signal \textbf{(a)} and the charge resistance \textbf{(b)} as a function of the trench height $A$ at $T=4.2$~K. The experimental data and the theory results are shown as points and lines, respectively. The shaded region represents the uncertainty due to device-to-device variation.
		The extended model is the same as described in the main text (green line and blue-shaded region). Here we include a simplified model only considering the increase in channel length due to the curved geometry (black line and red-shaded region).}
\end{figure}

The comparison between the experimental results and both the extended and the simplified model are shown in Fig.~\ref{fig:LTsimple}. We observe that this simplified model, which only considers the increase in total distance $L^{\prime}$ measured along the arclength of the film, fails to reproduce the strong suppression of $\Delta R_{\text{NL}}$ and the steep increase of $R$ with increasing trench height $A$. In comparison to the charge current, the spin current passes relatively unimpeded through the bends in the trench. 

%\clearpage

\end{document}